# Disparate exciton-phonon couplings for zone center and boundary phonons in solid-state graphite

Xuefei Feng[1,Ɨ], Shawn Sallis[1,Ɨ], Yu-Cheng Shao[1,2], Ruimin Qiao[1], Yi-Sheng Liu[1], Li Cheng Kao[1], Anton Tremsin[3], Zahid Hussain[4], Wanli Yang[1], Jinghua Guo[1], Yi-De Chuang[1]*

[1]*Advanced Light Source, Lawrence Berkeley National Laboratory, Berkeley, California 94720, USA*

[2]*Department of Physics, University of Houston, Houston, Texas 77204, USA*

[3]*Space Science Laboratory, University of California, Berkeley, California 94720, USA*

[4]*Materials Science Division, Lawrence Berkeley National Laboratory, Berkeley, California 94720, USA*

(Date: Jan. 2nd, 2020)

Ɨ: These authors contribute equally to this work.

*ychuang@lbl.gov

The exciton-phonon coupling in highly oriented pyrolytic graphite is studied using resonant inelastic X-ray scattering (RIXS) spectroscopy. With ~ 70 meV energy resolution, multiple low energy excitations associated with coupling to phonons can be clearly resolved in RIXS spectra. Using resonance dependence and the closed form for RIXS cross-section without considering the intermediate state mixing of phonon modes, the dimensionless coupling constant $g$ is determined to be 5 and 0.4, corresponding to the coupling strength of 0.42 eV +/- 40 meV and 0.21 eV +/- 30 meV, for zone center and boundary phonons respectively. The reduced *g* value for zone-boundary phonon may be related to its double resonance nature.

Carbon based materials, such as graphite, graphene, diamond, fullerene, and carbon nanotubes, have attracted much attention due to their unique electric and thermal properties that can be used in next generation electronic devices.[1,2] However, the lack of detailed understanding on the nature of interactions in these materials limits the ability to establish the relationship between their electronic structures and the prospective device performance. For example, the electron-phonon coupling (EPC), which determines material properties such as thermal and electric conductivities,[3] is one of the most important interactions in condensed matter systems.[4,5] The EPC also plays a critical role in the emergent phenomena like high $T_C$ superconductivity and its competing density wave states.[6] Although techniques such as optical Raman, electron energy loss (EELS), and angle-resolved photoemission (ARPES) spectroscopy have been used to study phenomena related to the EPC in carbon based materials, key parameter such as the coupling constant remains to be experimentally determined.[7]

Resonant inelastic X-ray scattering (RIXS) spectroscopy has recently been shown as a promising technique for studying the EPC in correlated materials.[8-13] In the RIXS process, the presence of core-hole and the photoexcited electron in the unoccupied state will change the local configuration around the interacting site.[14,15] The photoexcited electron can then interact with lattice vibrations during its decay to fill in the core-hole, as illustrated in the Frank-Condon picture in Figure 1(a), making it possible to investigate the EPC through the induced energy loss features in the RIXS spectra.[16,17] Complementary to optical Raman spectroscopy, high-resolution EELS, and ARPES,[18-20] RIXS is advantageous in terms of the unique sensitivity to the probed element and its chemical, orbital, and bonding states. In addition, the core-hole lifetime can be reduced by detuning the excitation photon energy away from the resonance,[21] effectively changing the probing time scale. With properly arranged measurement geometry to vary the photon momentum-transfer and its projection on the sample, one can obtain the momentum-resolved EPC with this technique.[22,23]

Clear phonon overtones in the RIXS spectra have been seen mainly in the molecular systems, and their finite size further permits the detailed analysis on the potential landscape and Rydberg levels for explaining these overtones.[24-29] However, such approach becomes a daunting task if not impossible for correlated materials. In fact, clear phonon excitations are rarely seen in the RIXS spectra of correlated materials except in the ones with strong EPC;[8,10] but even for those materials, different overtones never get cleanly resolved. In that regard, the current RIXS studies on highly oriented pyrolytic graphite (HOPG) can serve as a pedagogical example between these two limits.

RIXS studies on the EPC of carbon based materials have been limited by the poor energy resolution around 0.2 eV at C K-edge, which is too coarse to resolve the phonon excitations.[30,31] Here we revisit this topic with an improved energy resolution. The X-ray absorption (XAS) and RIXS experiments were carried out on the *ex-situ* cleaved, commercial HOPG at both iRIXS and qRIXS endstations at beamline 8.0.1 at the Advanced Light Source (ALS), Lawrence Berkeley National Laboratory (LBNL). [31-34] During the measurements, the π-polarization was used with Γ-K direction oriented in the horizontal scattering plane, see Figure 3(a) for the experimental geometry. For more details about the experiments, see Supplementary Information (SI).

The C K-edge XAS spectrum shown in Figure 1(b) was collected in the total electron yield (TEY) mode with the sample surface normal at 55° relative to the incident X-ray beam. The sharp features at 285.4 eV and 291.8 eV can be assigned to the excitations from C 1s core level to the unoccupied $\pi^*$ and $\sigma^*$ states, respectively.[35] The intensity ratio between these two features will depend on the sample geometry (see SI) and can exhibit spatial variation due to the ripples on the sample surface after cleavage; nevertheless, this ratio can be used to calibrate the sample position between temperature-dependent measurements during which the sample can move vertically from the cryostat expansion and contraction.

The RIXS map recorded at the same geometry with the X-ray spectrometer placed at 90° scattering angle is shown in Figure 1(c). The dominant features with emission energies between 270 eV and 280 eV are from the band transitions.[14] Zooming in the elastic peak region (zero energy loss, the white diagonal line in the figure), one can see the broadening when the excitation photon energies are tuned to C 1s → $\pi^*$ and $\sigma^*$ resonances. The strong dispersive shoulder next to the elastic peak around $\pi^*$ resonance (285.4 eV) can be attributed to the transition from the occupied $\sigma$-band at $\Gamma$ point to the C 1s core level, whose dispersion is manifested by the energy and momentum conservation with respect to the band structure.[12] On the other hand, the broadening of the elastic peak around $\sigma^*$ resonance (291.8 eV, yellow box) is related to the presence of phonon excitations that is the main focus of this work.

With improved energy resolution down to ~ 70 meV [34] and a proper arrangement of sample and spectrometer angles (the values are listed in each figure) to optimize the relative intensity between the elastic peak and the low energy excitations, the resulting photon energy dependent RIXS spectra at room temperature are shown in Figure 2(a). From these spectra, one sees that a small hump around 200 meV energy loss (pink curve, 289.80 eV) becomes a dominant feature when the excitation photon energy is tuned to σ* resonance (black curve, 291.80 eV). In the meantime, other broad humps at higher energy loss up to ~ 1 eV gain intensity such that 6 features can be clearly identified in the spectra (blue, red, and black curves at 291.40 eV, 291.65 eV, and 291.80 eV, respectively). These features, *tentatively* been regarded as the overtones of the fundamental at ~ 200 meV energy loss, may be explained by the Frank-Condon picture in Figure 1(a). The decrease in their intensities with respect to increasing the detuning $\Omega$ ($=\hbar(\omega_{in}-\omega_{res})$) from the resonance ($\hbar\omega_{res}$) can also be understood as the core-hole lifetime becomes much shorter that suppresses the RIXS channel involving the lattice degree of freedom.[21]

For quantitative analysis, we fit these low energy excitations in the 298 K, 291.8 eV spectrum (markers, Figure 2(b)) with Voigt functions (thin lines). The obtained peak energy position and the full-width-half-maximum (FWHM) are summarized in Figures 2(c) and 2(e), respectively. Although the peak position versus peak number seems to display a linear relationship, a close inspection shows that there are subtle variations. The positions of peak #2 to #6 consistently deviate from the linear extrapolation between the elastic peak (#0) and the first peak (#1), and the deviation exceeds the shaded region that denotes the possible peak energies when taking the error bar of peak #1 into consideration. The subtlety becomes evident if plotting the energy difference between the adjacent peaks in Figure 2(d): the oscillatory behavior is visible and is too large to be accounted for by the fitting errors. This variation implies that the energy of peak #n (n = integer) is not the multiple of the first peak at 190 ± 7 meV; instead, it is shifted towards lower energy loss with non-uniform offset. In addition, the FWHM of peak #2 is larger compared with the first and even the third peak, and the magnitude cannot be explained by the fitting error either.

We repeated the measurement at 80 K and the fitting results (bottom panel in Figure 2(b)) are shown as red open circles in Figures 2(c) – 2(f). Besides the enhanced elastic peak at 80 K, the intensity ratio between the first and the other peaks, as well as the peak energy position, are all temperature-independent even in the energy difference plot in Figure 2(c). The only discrepancy we can see besides the elastic peak intensity is the FWHM of peak #2, which is slightly narrower at 80 K. The agreement between the 298 K and 80 K data indicates that the observed anomalies in Figures 2(c) – 2(e) are intrinsic and unrelated to the fitting procedure. We also vary the sample and spectrometer angles slightly to see how these peaks change with respect to the experimental geometry (Figure 3(a)). Besides the strong response in the elastic peak (see Figures 3(b)-3(d)), subtle variations can only be seen after fitting these peaks with Voigt functions (colored lines). For comparison, we also overlay the 298 K data from Figures 2(c) and 2(f) in these figures as a reference (black markers). From these figures, we conclude that the peak energy positions do not depend on the experimental geometry within the fitting error (Figure 3(e)), although the relative peak intensities exhibit a weak geometry dependence.

With Einstein phonons, the RIXS cross-section can be calculated analytically to yield a closed form:[11,36-38]

$$I_{ph} \propto \sum_{n'=0}^{\infty} \left| \sum_{n=0}^{n'} \frac{B_{n'n}(g)B_{n0}(g)}{\Omega+i\Gamma+(g-n)\omega_0} + \sum_{n=n'+1}^{\infty} \frac{B_{nn'}(g)B_{n0}(g)}{\Omega+i\Gamma+(g-n)\omega_0} \right|^2 \cdot \delta(\omega-n'\omega_0) \quad (1)$$

In this equation, $0$, $n$, and $n'$ are the ground, intermediate, and final state phonon occupation number, respectively; $g$ is the dimensionless EPC, $\Gamma$ is the inverse core-hole lifetime, $\omega_0$ is the phonon mode energy, and $\Omega$ is the detuning of excitation photon energy from the resonance. $B_{ab}(g)$ in the numerator is the Frank-Condon factor. This equation has been used to calculate the $g$ value in correlated materials by comparing the intensities of phonon overtones (needs minimum the first and second order peaks) in the RIXS spectra at fixed excitation photon energies. However, by following this approach, one inherently assumes that the peaks #2 to #5 in Figures 2(b), 3(b) – 3(d) are the real overtones of the first peak with their energy positions at the multiple of 190 ± 7 meV. This assumption does not agree with the anomalies seen in the peak positions (Figures 2(c) and 2(d)), the FWHM (Figure 2(e)), and the relative intensity variations in Figure 3(f) where the results suggest the presence of different phonon modes besides the overtones in the RIXS spectra.

Soft X-ray photons typically do not carry large enough momentum. For 291.80 eV photon and the spectrometer at 77.5° (Figure 2(a), the angles are defined in Figure 3(a)), the photon momentum transfer $\Delta q$ is around 0.23 A$^{-1}$, which is ~ 13% of the Brillouin Zone size along the Γ-K direction. At nearly specular geometry, the in-plane projection $\Delta q_{//}$ can be even smaller: only <3% of the Γ-K size for the geometry shown in Figure 2. Because of this, the conservation of momentum requires the electron coupling to phonon mode(s) with very small net momentum. When the excitation photon energy is tuned to the resonance of σ* excitonic state, the transition occurs at the bottom of the σ* band at Γ point.[15] Since we are effectively looking at the "optical transition" at the C 1s → σ* resonance, we can resort to the results of optical Raman spectroscopy to identify the phonon modes seen in the RIXS spectra.[39]

In the optical Raman spectroscopy, there are two prominent phonon modes with $E_{2g}$ (G mode at Γ point, 1583 cm$^{-1}$ or 196 meV) and $A_1$ symmetry (D mode at K point; it becomes Raman active through the double resonance so that its energy is 2700 cm$^{-1}$ or 335 meV), respectively.[40] The energy of these modes agrees very well with the energy position of peak #1 and #2: peak #1 at 190 ± 7 meV can be assigned to the G mode whereas peak #2 at 335 ± 7 meV can be assigned to the 2D mode. Interestingly, the oscillatory behavior in Figure 2(d) suggests that the energy of peak #3 will be ~ 190 meV higher than peak #2, prompting us to assign it to G+2D mode which is also seen in the Raman spectroscopy. The assignment of prominent energy loss features in Figure 2(b) to different phonon modes plus the enhanced peak #2 (2D) and #3 (G+2D) that preclude the separation of overtones like 2G and 3G from them prohibit us from using eq. (1) to directly fit the intensities of fundamental and high order overtones to extract the *g* value.[10-12] Alternatively, one can use the resonance effect to determine the EPC as recently demonstrated by Rossi *et al*.[11] However, the mixing of G and D modes in the virtual intermediate state in the RIXS process can complicate the analysis and lead to a sizeable uncertainty in the coupling constant *g*.[22] Despite lacking the detailed theoretical calculations to account for this mixing effect, following the treatment by Rossi *et al.* without considering the mixing effect (e.g. attributing the peak #1 solely to the G mode from the longitudinal and transverse optical branches) may still yield valuable insight to the nature of the excitonic coupling to these phonons and inspire further theoretical studies.

To proceed with this approach, the self-absorption effect needs to be corrected before one can compare the spectral intensities recorded at different excitation photon energies. For more details about this correction, see SI. The intensity of G mode obtained from a Voigt function fitting is shown as the blue circles in Figure 4. In the same figure, we also overlay the XAS spectrum marked by the gray shaded area. Unlike in the case of cuprates, the intensity of this phonon remains appreciable even when the excitation photon energy is detuned by 1.0 eV, suggesting a strong coupling nature. The intensity for exciting a single phonon can be expressed as (*n'* = 1):

$$I_{ph} \propto \frac{e^{-2g}}{g} \left| \sum_{n=0}^{\infty} \frac{g^n (n-g)}{n![\Omega + i\Gamma + (g-n)\omega_0]} \right|^2 \quad (2)$$

We use this equation to calculate the intensity as a function of detuning $\Omega$ for several *g* values (blue curves) and *g* = 5 gives the best agreement. This *g* value corresponds to the EPC strength $M_G = \sqrt{g}\omega_0$ of ~ 0.42 eV +/- 40 meV.

Assuming the *g* value does not change with the excitation photon energy, we can then calculate the contribution of this mode for fundamental up to 5$^{th}$ order overtone using eq. (1) (*n'*=1 ~ 5). For each order, we also broaden the feature with a linearly increased width

(15%) as suggested by Figure 2(e). The sum is then subtracted from the self-absorption corrected RIXS spectra to yield the fundamental component of 2D mode, whose intensity is shown as the green squares in Figure 4. Contrasting to the behavior of G mode, the 2D mode intensity is quickly suppressed with increasing the detuning $\Omega$, indicating a much weaker coupling. By plotting the eq. (1) with several $g$ values (green curves), one sees that $g \sim 0.4$ gives a good agreement. This value gives the EPC strength $M_{2D}$ of $\sim 0.21$ eV +/- 30 meV. We note that the approach of separating the estimated overtone of the G mode (2G) from 2D mode depends heavily on the assumption that the feature at $190 \pm 7$ meV comes solely from the G mode; and by including the intermediate state mixing between D and G modes, the aforementioned approach would underestimate (overestimate) the $g$ value for 2D (G) mode. Nevertheless, the values reported from RIXS for both phonon modes are much larger than that from recent time-resolved ARPES (TR-ARPES).[41]

In the RIXS process, the photoexcited electron and the poorly screened core-hole in the form of exciton at $\sigma^*$ resonance can exert a much larger perturbation to the surrounding C atoms than the particle-hole pair created in the optical Raman process; therefore, it is plausible to see much stronger 2D and G+2D modes relative to the G mode in RIXS (relative intensity of 0.75 and 0.5 respectively) than in Raman (0.27 and 0.02). With such large perturbation, the strongly coupled phonon modes like G and 2D are prime candidates to be excited due to the unique HOPG phonon band structure at $\Gamma$ and K points with Kohn anomaly.[42-44] The large $g$ values for G and 2D modes relative to that from TR-ARPES can be attributed to the presence of $\sigma^*$ exciton that is absent in the ARPES process. Furthermore, we conjecture that since this 2D mode involves double resonance to become

Raman active (also for RIXS at such low photon energy), the scattering cross-section will be suppressed once core-hole lifetime is reduced, thus giving a smaller coupling. Our work demonstrates that the improved RIXS spectral resolution coupled to versatile tuning knobs like resonance, experimental geometry, etc. can facilitate the determination and differentiation of phonon excitations in carbon-based materials, shedding light on the nature of phononic interactions that can be useful for carbon-based energy applications.[45,46]

We thank Dr. Keith Gilmore for the insightful discussion. This work at the advanced Light Source is supported by the Director, Office of Science, Office of Basic Energy Sciences, of the U.S. Department of Energy (DOE) under Contract No. DE-AC02-05CH11231. X.F. acknowledges the support of Joint Center for Energy Storage Research (JCESR), an Energy Innovation Hub funded by U.S. DOE, office of Science, Basic Energy Sciences.

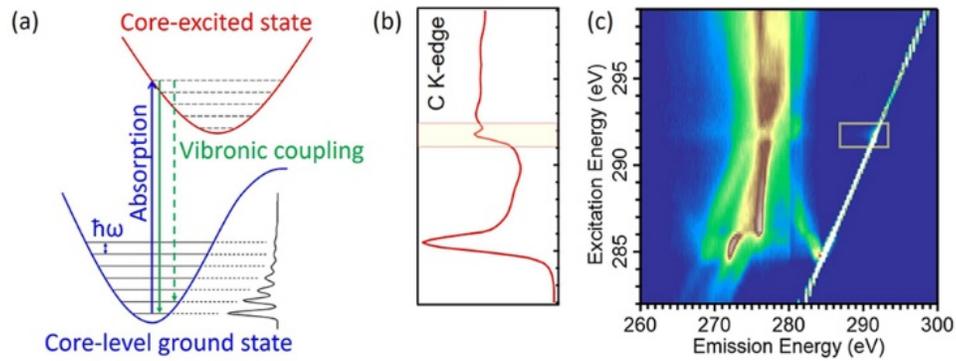

Figure 1. (a) Schematic illustration of Frank-Condon picture showing the transition from vibrationally excited states to the ground state. (b) Total electron yield mode of XAS spectrum at C K-edge recorded with sample surface normal at 55° angle at room temperature. (c) RIXS map of HOPG at room temperature. The broadening of the elastic peak is observed at excitation energies between 291 eV and 292 eV (light yellow box).

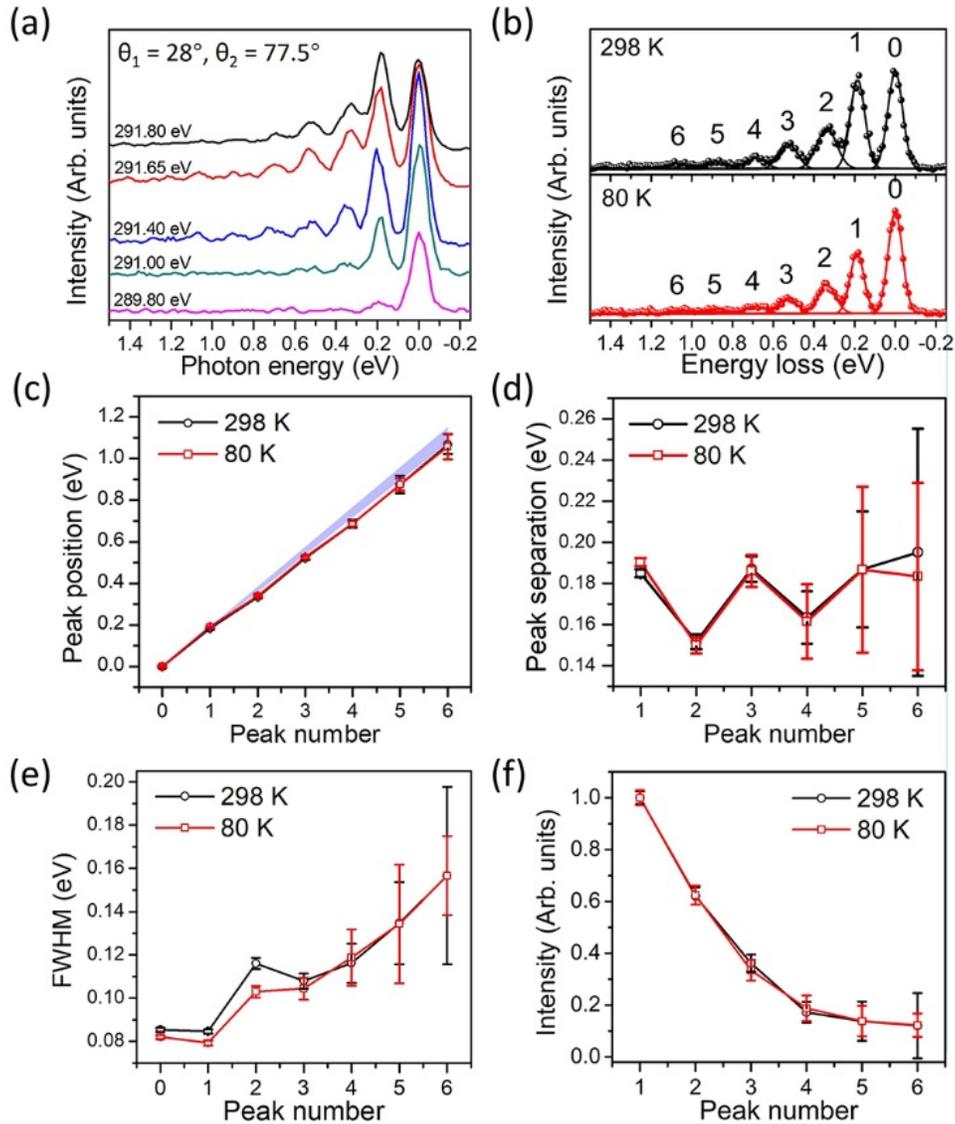

Figure 2. (a) Room temperature RIXS spectra recorded with excitation energies around the σ* resonance (291.80 eV). The inset shows the experimental geometry. (b) RIXS spectra recorded at 298 K (top, black markers) and 80 K (bottom, red markers) with geometry shown in the inset of Figure 2(a). The excitation photon energy is 291.80 eV. The lines are the Voigt function fitting. The elastic peak and phonons are labeled as Peak #0 and Peaks #1~#6, respectively. (c) Energy position of the peaks with the elastic peak set to 0 eV. The blue shaded area shows the possible energy position by connecting the peak #0 to the maximum/minimum of peak #1 (the error bar). (d) The energy difference between the adjacent peaks from panel (c). (e) Full-width-half-maximum (FWHM) of the peaks from the fitting shown in panel (b). (f) Comparison of the intensities of peaks at 80 K (red) and 298 K (black). The intensity of peak #1 is set to 1.0.

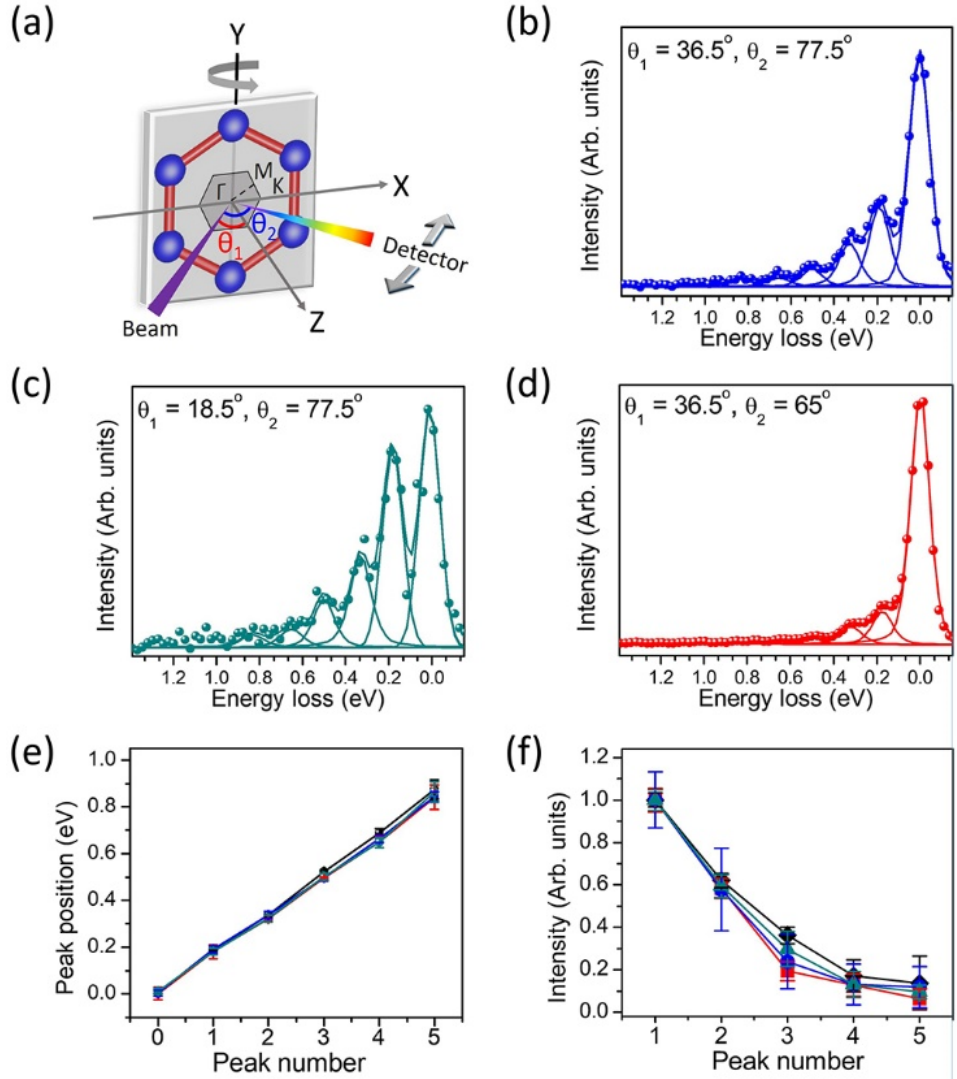

Figure 3. (a) Schematic diagram of experimental geometry. (b)-(d) Room temperature RIXS spectra with excitation photon energy 291.80 eV and experimental geometry shown in panel (a). The sample and spectrometer angles are listed in each figure. (e) The peak positions from fitting spectra in panels (b)-(d) with Voigt functions. (f) The normalized intensities of the peaks from the fitting. The intensity of first peak is set to 1.0. The symbol color scheme in panels (e) and (f) adopts the trace color in panels (b)-(d). The black symbols in (e) and (f) are from Figures 2(c) and 2(f) as reference.

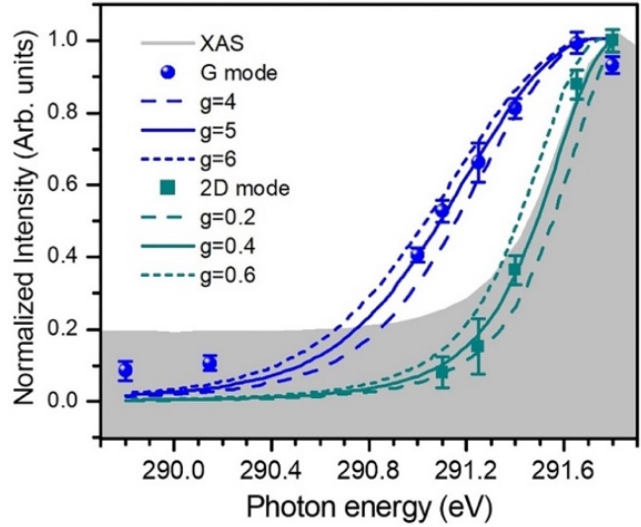

Figure 4. Overlay of normalized XAS spectrum (shaded area), $E_{2g}$ (G mode) peak intensity (filled blue circles), $2*A_{1g}$ (2D mode) peak intensity (filled green squares) and calculated intensity using eq. (2) with different $g$ value (solid lines). The parameters $\omega_0 = 0.190$ eV for G mode, $\omega_0 = 0.335$ eV for 2D mode, $\Gamma = 0.414$ eV (~10 fs) are used in calculations.

# Supplementary Information for

# "Disparate exciton-phonon couplings for zone center and boundary phonons in solid-state graphite"

Xuefei Feng[1,Ɨ], Shawn Sallis[1,Ɨ], Yu-Cheng Shao[1,2], Ruimin Qiao[1], Yi-Sheng Liu[1], Li Cheng Kao[1], Anton Tremsin[3], Zahid Hussain[4], Wanli Yang[1], Jinghua Guo[1], Yi-De Chuang[1]*

[1]*Advanced Light Source, Lawrence Berkeley National Laboratory, Berkeley, California 94720, USA*

[2]*Department of Physics, University of Houston, Houston, Texas 77204, USA*

[3]*Space Science Laboratory, University of California, Berkeley, California 94720, USA*

[4]*Materials Science Division, Lawrence Berkeley National Laboratory, Berkeley, California 94720, USA*

Ɨ: These authors contribute equally to this work.
*ychuang@lbl.gov

(Date: Jan. 2nd, 2020)

**S1. Experimental details**

The highly oriented pyrolytic graphite (HOPG) used in the experiments was purchased from SPI supplies (Grade SPI-1, Item 498HP-AB). The sample was cleaved *ex-situ* to yield a fresh surface before introduced into a high vacuum experimental chamber for measurements (base pressure better than $5*10^{-9}$ torr). The Γ-K direction was oriented roughly in the horizontal scattering plane with π-polarization used during the measurements, see Figure 3(a) for the schematic illustration of experimental geometry in the manuscript. The sample mosaicity of ~ ±5° was determined from Laue diffraction.

The X-ray absorption (XAS) and RIXS experiments were carried out at both iRIXS and qRIXS endstations at beamline 8.0.1 at the Advanced Light Source (ALS), Lawrence Berkeley National Laboratory (LBNL). XAS spectrum was collected in total electron yield (TEY) mode, through monitoring the sample-to-ground drain current, and normalized using the photocurrent from an upstream Au mesh. The RIXS spectra were recorded using the modular X-ray spectrometers installed in these two endstations. The RIXS map shown in Figure 1(c) of the manuscript was obtained at iRIXS endstation (BL8.0.1.1) with the spectrometer placed at 90° scattering angle relative to the incident X-ray beam. A commercial CCD detector having ~ 27 μm effective pixel resolution was used in this spectrometer. With 50 μm (entrance slit) × 50 μm (exit slit) beamline slit setting, the combined resolving power (E/ΔE, beamline plus spectrometer) was around 1500.

The higher resolution RIXS spectra used in Figures 2 and 3 in the manuscript were recorded using the spectrometers at qRIXS endstation (BL8.0.1.3). A Timepix detector with 7.5 μm effective pixel resolution after centroiding algorithm was used for the measurements. The scattering angle (90°-$\theta_2$ per notation in Figure 3(a) in the manuscript) of spectrometers can be varied from 27.5° (forward scattering) to 152.5° (backward scattering), and the exact sample and spectrometer angles are listed in the figures. With 40 μm (entrance slit) × 20 μm (exit slit) beamline slit setting, the combined E/ΔE determined from the FWHM of elastic peak was around 4000. It took about 4 hours to acquire each spectrum with this resolving power.

**S2. Self-absorption correction for RIXS spectra**

The emitted X-rays from the interior of material can be reabsorbed by the material itself before leaving the surface, resulting in a distorted RIXS spectral line shape that depends on the energy of X-rays and the measurement geometry (see Figure S1(a)). This self-absorption effect needs be corrected before comparing the spectral intensity recorded with different excitation photon energies and experimental geometry. The self-absorption correction factor can be expressed as:[1-4]

$$C = \frac{1}{1+r \cdot u} \tag{1}$$

Here, $r$ is the ratio of absorption coefficients between the incident and emitted X-rays:

$$r = \frac{1+\frac{\alpha_2}{\alpha_0}}{1+\frac{\alpha_1}{\alpha_0}} \tag{2}$$

$\alpha_0$, $\alpha_1$, and $\alpha_2$ are the absorption coefficients before the resonance edge, at the incident and emitted X-ray energies, respectively (see Figure S1(b)). $u$ is the ratio of path lengths from the surface to a depth $d$ between the incident ($S_1$) and emitted X-rays ($S_2$) at angles $\theta_1$ and $\theta_2$:

$$u = \frac{S_2}{S_1} = \frac{d \cdot \sin\theta_1}{d \cdot \sin\theta_2} = \frac{\sin\theta_1}{\sin\theta_2} \tag{3}$$

For the spectra in Figure 2 of the manuscript, the angles are 62° and 40.5°, respectively. Taking the excitation photon energy of 291.80 eV as an example, the calculated $C$ as a function of energy loss is shown in Figure S1(c). The corrected spectrum shown as the red curve in Figure S1(d) is obtained by dividing the raw spectrum (black curve, normalized to the incident photon flux only) with this factor *C*. As one can see, the self-absorption correction mainly affects the intensities of the first three dominant energy loss features in the spectrum.

All spectra shown in Figure 2 and 3 of the manuscript have been corrected following this procedure.

**S3. Calculation of the intensity of phonon overtones**

**S3a. Estimate the EPC strength for G mode (E$_{2g}$ phonon)**

Using the self-absorption corrected RIXS spectra at different excitation photon energies, we can fit the first E$_{2g}$ phonon (G mode) feature with a Voigt function to obtain its spectral weight and derive the dimensionless EPC constant *g* as described in the manuscript. Figure S2(a) shows an example of such fitting where the open circles are the experimental data (self-absorption corrected) and the blue shaded area is the fitted G phonon component. The green curve is a Gaussian fitting of the elastic peak, and the red curve is the sum of Voigt functions for other high energy phonons.

The intensity of G mode as a function of excitation photon energy is shown as solid black circles in Figure S2(b). To estimate the dimensionless EPC constant *g*, we resort to the RIXS cross-section with Einstein phonon:[1, 5, 6]

$$I_{ph} \propto \frac{e^{-2g}}{g} \left| \sum_{n=0}^{\infty} \frac{g^n (n-g)}{n! [\Omega + i\Gamma + (g-n)\omega_0]} \right|^2 \qquad (4)$$

The phonon mode energy $\omega_0$ from our RIXS data is 190±7 meV, which agrees with previous Raman results (~1583 cm$^{-1}$ or 196 meV). The core-hole lifetime $\Gamma$ for 1s -> $\sigma^*$ core exciton state is set to ~10 fs (414 meV).[7] Calculated intensity using eq. (4) with several $g$ values are shown as color lines in Figure S2(b). The summation series converges when $n$ is larger than 25, so we compute the sum up to $n = 29$. Based on this figure, the $g = 5$ curve gives the best agreement with the experimental data. With this $g$ value, the EPC strength $M_G = \sqrt{g}\omega_0$ can be evaluated to be 0.42 eV +/- 40 meV.

**S3b. Estimate the EPC strength for 2D mode (2*A$_{1g}$ phonon)**

Following Ament et al., the RIXS cross-section with Einstein phonon can be expressed as:[5]

$$\frac{d^2\sigma}{d\Omega d\omega} = N|T_{el}(\varepsilon',\varepsilon)|^2 \sum_{n'=0}^{\infty} \left| \sum_{n=0}^{n'} \frac{B_{n_i'n}(g)B_{n0}(g)}{z+(g-n)\omega_0} + \sum_{n=n'+1}^{\infty} \frac{B_{nn_i'}(g)B_{n0}(g)}{z+(g-n)\omega_0} \right|^2 \delta(\omega - n'\omega_0) \quad (5)$$

Here, 0, $n$, and $n'$ are the ground, intermediate, and final state phonon occupation number, respectively. $B_{ab}(g)$ is the Franck-Condon factor:

$$B_{ab}(g) = \sqrt{e^{-g}a!b!} \sum_{l=0}^{b} \frac{(-1)^a(-g)^l \sqrt{g}^{a-b}}{(b-l)!l!(a-b+l)!} \quad (6)$$

$z = \hbar\omega_k - E_{res} + i\Gamma$ with resonance energy $E_{res}$ chosen at the maximum in XAS spectrum (291.80 eV). Note that eq. (4) is just the result with $n' = 1$. Using this equation, the closed-form solutions with different final state $n'$ (up to 5$^{th}$ order) are summarized in Table S1.

| n' | Closed-form solution |
|---|---|
| 0 | $I^{(0)} = N|Tel(\varepsilon',\varepsilon)|^2 \cdot e^{-2g} \cdot \left|\sum_{n=0}^{\infty} \frac{g^n}{n![z+(g-n)\omega_0]}\right|^2$ |
| 1 | $I^{(1)} = N|Tel(\varepsilon',\varepsilon)|^2 \cdot \frac{e^{-2g}}{g} \cdot \left|\sum_{n=0}^{\infty} \frac{g^n(n-g)}{n![z+(g-n)\omega_0]}\right|^2$ |
| 2 | $I^{(2)} = N|Tel(\varepsilon',\varepsilon)|^2 \cdot \frac{e^{-2g}}{2g^2} \cdot \left|\sum_{n=0}^{\infty} \frac{g^n\left[(n-g)^2-n\right]}{n![z+(g-n)\omega_0]}\right|^2$ |
| 3 | $I^{(3)} = N|Tel(\varepsilon',\varepsilon)|^2 \cdot \frac{e^{-2g}}{6g^3} \cdot \left|\sum_{n=0}^{\infty} \frac{g^n\left\{(n-g)^3-n[3(n-g-1)+1]\right\}}{n![z+(g-n)\omega_0]}\right|^2$ |
| 4 | $I^{(4)} = N|Tel(\varepsilon',\varepsilon)|^2 \cdot \frac{e^{-2g}}{24g^4} \cdot \left|\sum_{n=0}^{\infty} \frac{g^n\left\{(n-g)^4-n\left[6(n-g-1)^2+4(n-g-1)-3n+4\right]\right\}}{n![z+(g-n)\omega_0]}\right|^2$ |
| 5 | $I^{(5)} = N|Tel(\varepsilon',\varepsilon)|^2 \cdot \frac{e^{-2g}}{120g^5}$ $\cdot \left|\sum_{n=0}^{\infty} \frac{g^n\left\{(n-g)^5-n\left[10(n-g-1)^3-5(n-g-1)^2-15g(n-g-1)+10(n-g-1)+5g+1\right]\right\}}{n![z+(g-n)\omega_0]}\right|^2$ |

Table S1: Closed-form solutions of the intensity of a single phonon in RIXS process with different orders. The intensity of phonon overtones $I^{(n')}$ with different order number (n') are shown. Here $n' = 0$ is the elastic peak.

In Figure S3(a), we overlay the experimental RIXS spectrum at 291.80 eV (open circles) with the simulated one (red curve) obtained by calculating the G mode phonon contributions up to 5$^{th}$ order using the equations in Table S1. For high order phonons, their widths are increased linearly based on the results in Figure 2(e) of the manuscript for simplicity: the $n^{th}$ order G phonon will have a width of $(80 + (n-1)*20)$ meV. The shaded area in Figure S3(a) is the difference between the experimental and simulated spectra, which is taken as the spectral weight of the 1$^{st}$ order 2D mode. Its excitation photon energy dependence is shown as black filled circles in Figure S3(b). In the same figure, we also show the calculated intensity curves using eq. (4) with the 2D mode energy $\omega_0 = 335$ meV and $\Gamma = 414$ meV for several $g$ values. $g \sim 0.4$ gives the best overall agreement. With this $g$ value, the EPC strength for 2D mode can be estimated to be 0.21 eV +/- 30 meV.

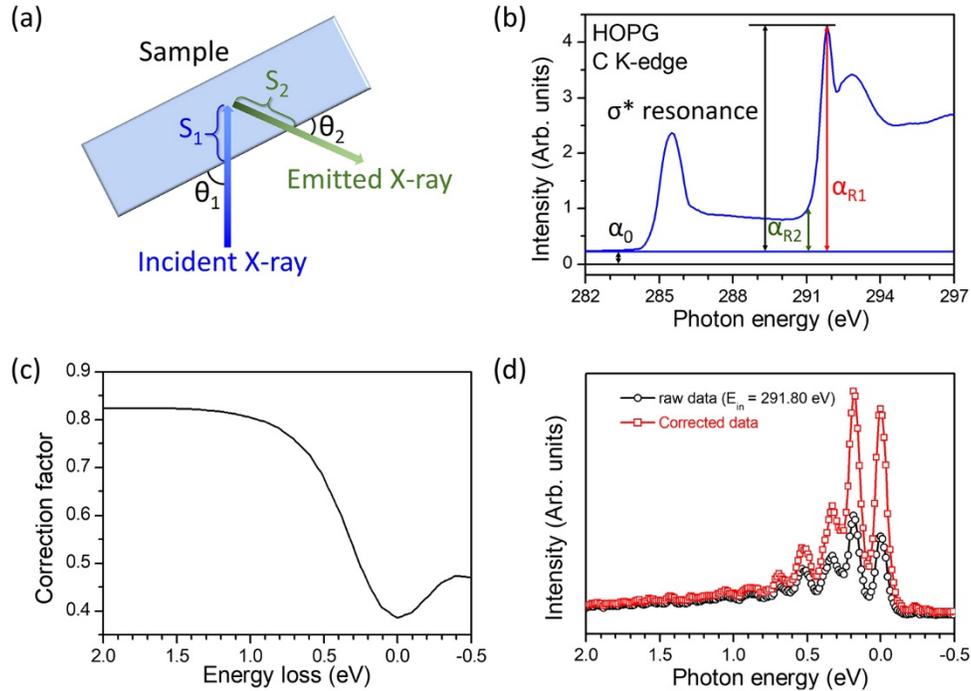

Figure S1: (a) Schematic illustration of the experimental geometry. $S_1$ and $S_2$ are the path lengths of incident and emitted X-rays from the sample surface to a depth d at $\theta_1$ and $\theta_2$ grazing incidence angles, respectively. (b) C K-edge XAS spectrum with sample surface normal at 28° relative to the incident X-ray beam. Taking 291.80 eV incident energy ($E_{in}$) as an example, $\alpha_0$, $\alpha_1$, and $\alpha_2$ are the intensities of the spectrum at pre-edge, incident, and emitted X-ray energies. Note that the relative intensity between $\sigma^*$ ad $\pi^*$ features is different from that shown in Figure 1(b) in the manuscript. The difference is caused by the sample geometry relative to the incident X-ray beam. (c) Self-absorption correction factor at the incident energy of 291.80 eV. (d) The comparison of self-absorption corrected (red open squares) and uncorrected (black open circles) spectrum with $E_{in}$ =291.80 eV.

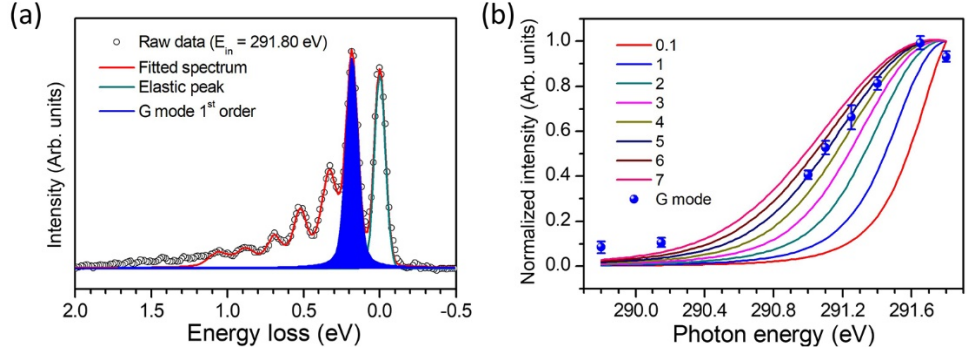

Figure S2: (a) Overlay of experimental (open circles) and fitted RIXS spectra (red curve) with incident photon energy of 291.80 eV. The fitted spectrum is a sum of Voigt functions for different phonons and the component for first order G phonon is shown as a blue shaded area. The green curve is a Gaussian fitting of the elastic peak to determine the energy resolution around 70 meV. (b) Spectral weight of the first order G mode as a function of excitation photon energy (black solid circles). The lines are the calculated intensity using eq. (4) with different coupling constant $g$. All curves are normalized to 1 at 291.80 eV.

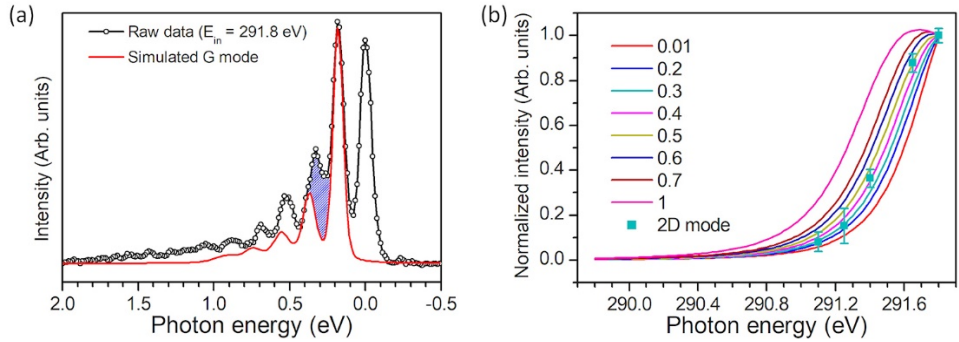

Figure S3: (a) Overlay of experimental (open circles) and simulated RIXS spectra (red curve) with incident photon energy of 291.80 eV. The simulation assumes a fixed $g = 5$ and includes the G phonon contributions up to $5^{th}$ order overtone calculated from the equations in Table S1. The width of high energy phonons increases linearly according to the results in Figure 2(e) of the manuscript. The shaded area is the intensity difference between the experimental and simulated spectra and is taken as the spectral weight of $1^{st}$ order 2D mode phonon. (b) Spectral weight of first order 2D mode as a function of excitation photon energy (black open circles). The lines are the calculated intensity using eq. (4) with different coupling constant $g$. The 2D mode energy is set to 335 meV. All curves are normalized to 1 at 291.80 eV.